\documentclass[%
prl,twocolumn,showpacs
 amsmath,amssymb,
 aps,
]{revtex4-1}

\usepackage{graphicx}
\usepackage{dcolumn}
\usepackage{bm}

\newcommand{\EQ}{\begin{equation}}
\newcommand{\EN}{\end{equation}}
\newcommand{\be}{\begin{equation}}
\newcommand{\ee}{\end{equation}}
\newcommand{\bea}{\begin{eqnarray}}
\newcommand{\eea}{\end{eqnarray}}

\begin{document}


\title{Vortex mass in the three-dimensional $O(2)$ scalar theory}

\author{Gesualdo Delfino$^{1,2}$, Walter Selke$^3$ and Alessio Squarcini$^{4,5}$}
\affiliation{$^1$SISSA -- Via Bonomea 265, 34136 Trieste, Italy\\
$^2$INFN sezione di Trieste, 34100 Trieste, Italy\\
$^3$Institute for Theoretical Physics, RWTH Aachen University, 52056 Aachen, Germany\\
$^4$Max-Planck-Institut f\"ur Intelligente Systeme, Heisenbergstr. 3, D-70569, Stuttgart, Germany\\
$^5$IV. Institut f\"ur Theoretische Physik, Universit\"at Stuttgart, Pfaffenwaldring 57, D-70569 Stuttgart, Germany}


\begin{abstract}
We study the spontaneously broken phase of the $XY$ model in three dimensions, with boundary conditions enforcing the presence of a vortex line. Comparing Monte Carlo and field theoretic determinations of the magnetization and energy density profiles, we numerically determine the mass of the vortex particle in the underlying $O(2)$-invariant quantum field theory. The result shows, in particular, that the obstruction posed by Derrick's theorem to the existence of stable topological particles in scalar theories in more than two dimensions does not in general persist beyond the classical level. 
\end{abstract}

\maketitle




Topological excitations are among the most fascinating objects in quantum field theory (QFT) \cite{Coleman,Weinberg}. Being associated to extended configurations of the fields appearing in the action, they are intrinsically non-perturbative and difficult to characterize as quantum particles. A well known exception is provided by two-dimensional space-time, where sine-Gordon solitons correspond, through fermionization, to the fundamental fields of the massive Thirring model \cite{Coleman_SG}; in addition, integrability allows a full and exact quantum description \cite{ZZ}. No similar methods, however, are available in higher dimensions.

In three dimensions the simplest theory with symmetry properties allowing for topological excitations -- vortices -- is the $O(2)$-invariant scalar theory. This describes the universality class of the $XY$ lattice model, which includes the superfluid transition of ${}^4$He as a particularly interesting representative (see \cite{LSNCI}). While a direct responsibility of vortices in the phase transition of the three-dimensional $O(2)$ theory has been debated \cite{KSW,Weichman}, it is a fact that the transition is of the type associated to spontaneous symmetry breaking, which occurs also in absence of non-trivial topology. In field theory, vortices in the scalar theory have usually been considered only to point a problem, i.e. that the energy (mass) of the static classical solution diverges logarithmically \cite{Ryder}, a particular case of Derrick's theorem \cite{Derrick,Coleman,Weinberg}. This divergence at the classical level suggested the absence of a vortex particle in the scalar QFT.

In this paper we consider the three-dimensional $XY$ model in its spontaneously broken phase, slightly below the critical temperature $T_c$, with boundary conditions enforcing the presence of a vortex line. As the other properties of the near-critical system, the corresponding energy density and order parameter profiles have to be accounted for by the $O(2)$ scalar QFT describing the continuum limit. Remarkably, these profiles are calculable in the field-theoretical framework, and we compare the analytic results with the numerical determination obtained by Monte Carlo simulations, finding excellent agreement as we vary the temperature and the end-to-end distance of the vortex line. In the process we numerically determine the mass $m_V$ of the vortex particle, which for small $|T-T_c|$ can be expressed as 
\EQ
m_V\approx 2.1\,m_+\,,
\label{ratio}
\EN
where $m_+$ is the mass of the fundamental particles in the phase with unbroken symmetry ($T>T_c$). This result provides the first direct verification that Derrick's theorem, as a statement concerning classical field configurations, does not prevent the existence of stable topological particles in quantum theories of self-interacting scalar fields in more than two dimensions.

We consider the $XY$ model with reduced Hamiltonian 
\EQ
{\cal H}=-\frac{1}{T}\sum_{<i,j>}{\bf s}_i\cdot{\bf s}_j\,,
\label{H}
\EN
where ${\bf s}_i$ is a two-component unit vector (spin) located at the site $i$ of a cubic lattice, and the sum is performed over all pairs of nearest neighboring sites. We focus on the case $T<T_c$ in which the $O(2)$ symmetry of the Hamiltonian is spontaneously broken, i.e. $\langle{\bf s}_i\rangle\neq 0$; $\langle\cdots\rangle$ denotes the average over spin configurations weighted by $e^{-{\cal H}}$. Close to $T_c$ the intrinsic length scale of the system becomes much larger than lattice spacing and the system is described by a $O(2)$-invariant Euclidean scalar field theory, which in turn is the continuation to imaginary time of a QFT in $(2+1)$ dimensions. Switching to notations of the continuum, we denote by $x=(x_1,x_2,\tau)\equiv({\bf x},\tau)$ a point in Euclidean space, $\tau$ being the imaginary time direction, and by ${\bf s}(x)=(s_1(x),s_2(x))$ the two-component spin field. The field theory is the usual one defined by the action
\EQ
{\cal A}=\int d^3x\left\{[\partial_\mu{\bf s}(x)]^2+g_2\,{\bf s}^2(x)+g_4[{\bf s}^2(x)]^2\right\}\,,
\EN
with the $XY$ critical point reachable tuning the couplings (see e.g. \cite{Z-J}).

\begin{figure}[t]
\begin{center}
\includegraphics[width=6cm]{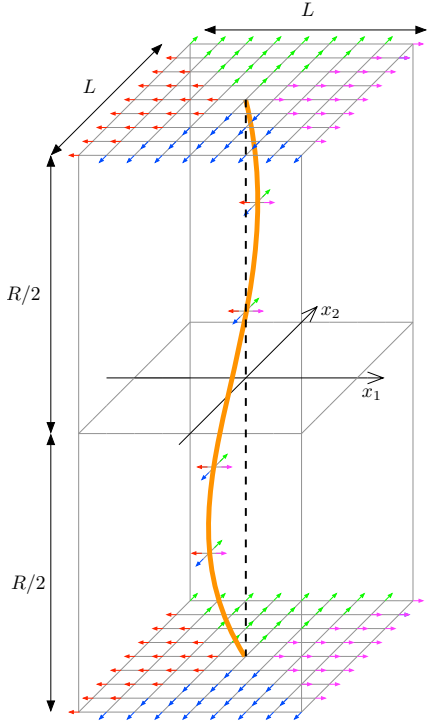}
\caption{Geometry considered for the $XY$ model below $T_c$. Boundary spins point outwards orthogonally to the vertical external surfaces. On the top and bottom surfaces they are fixed as indicated, so that a vortex line (one configuration is shown) runs between the central points of these surfaces.}
\label{geometry}
\end{center}
\end{figure}

We consider the system as defined in the volume $x_1\in(-L/2,L/2)$, $x_2\in(-L/2,L/2)$, $\tau\in(-R/2,R/2)$, with $L\to\infty$ and $R$ large but finite. The boundary conditions are chosen in such a way that all spins on the external surfaces parallel to the $\tau$-axis point outwards orthogonally to the surface. This implies the formation of a vortex on each section with constant $\tau$, with the vortex center forming a vortex (or defect) line as $\tau$ varies. Boundary conditions on the surfaces $\tau=\pm R/2$ are fixed as shown in figure~\ref{geometry}, so that the endpoints of the vortex line are fixed at ${\bf x}=0$, $\tau=\pm R/2$. The vortex line corresponds to the trajectory in imaginary time of a topological particle (the vortex $V$) in the $(2+1)$-dimensional QFT \cite{vortex}. 

The boundary conditions at $\tau=\pm R/2$ act as boundary states $|B(\pm R/2)\rangle=e^{\pm\frac{R}{2}H}|B(0)\rangle$ of the Euclidean time evolution; here $H$ denotes the Hamiltonian of the quantum system. These boundary states can be expanded on the basis of asymptotic particle states of the QFT, and will contain the vortex as the contribution with minimal energy, i.e.
\EQ
|B(\pm R/2)\rangle=\int\frac{d{\bf p}}{(2\pi)^2E_{\bf p}}\,a_{\bf p}\,e^{\pm\frac{R}{2}E_{\bf p}}\,|V({\bf p})\rangle+\ldots\,,
\label{B}
\EN
where ${\bf p}$ is the two-component momentum of the particle, $E_{\bf p}=\sqrt{{\bf p}^2+m_V^2}$ its energy, $a_{\bf p}$ an amplitude, and we normalize the states by $\langle V({\bf p}')|V({\bf p})\rangle=(2\pi)^2E_{\bf p}\,\delta({\bf p}-{\bf p}')$. In the calculations performed with the boundary conditions we have chosen (which we indicate with a subscript ${\cal B}$) the one-vortex contribution in (\ref{B}) determines the asymptotics for $R\gg 1/m_V$ (indicated below by the symbol $\sim$). Then we have 
\bea
Z_{\cal B} & \equiv & \langle B(R/2)|B(-R/2)\rangle=\langle B(0)|e^{-RH}|B(0)\rangle
\label{Z}\\
&\sim & |a_0|^2\int\frac{d{\bf p}}{(2\pi)^2 m_V}\,e^{-(m_V+\frac{{\bf p}^2}{2m_V})R}=\frac{|a_0|^2}{2\pi R}\,e^{-m_VR},
\nonumber
\eea
while for the expectation value of a field $\Phi$ we obtain
\bea
& \langle\Phi({\bf x},0)\rangle_{\cal B}=\frac{1}{Z_{\cal B}}\,\langle B(R/2)|\Phi({\bf x},0)|B(-R/2)\rangle \hspace{1.8cm}
\label{vPhi0}\\
&\sim \frac{R}{(2\pi)^3 m_V^2}\int d{\bf p}_1d{\bf p}_2\,F_\Phi({\bf p}_1|{\bf p}_2)\,e^{-\frac{R}{4m_V}({\bf p}_1^2+{\bf p}_2^2)+i{\bf x}\cdot({\bf p}_1-{\bf p}_2)},
\nonumber
\eea
where
\EQ
F_\Phi({\bf p}_1|{\bf p}_2)=\langle V({\bf p}_1)|\Phi(0,0)|V({\bf p}_2)\rangle\,,\hspace{.5cm}{\bf p}_1,{\bf p}_2\to 0
\label{ff}
\EN
is the low-energy limit of the form factor of the field on the vortex state. Its behavior determines the final form of (\ref{vPhi0}). 

The expectation value $\langle{\bf s}({\bf x},0)\rangle_{\cal B}$ (magnetization) has to interpolate between zero at ${\bf x}=0$ (where the symmetry is unbroken) and the asymptotic value
\EQ
\lim_{|{\bf x}|\to\infty}\langle{\bf s}({\bf x},0)\rangle_{\cal B}= v\,\hat{\bf x}\,,
\label{c1}
\EN
where $\hat{\bf x}={\bf x}/|{\bf x}|$, and $v=|\langle{\bf s}({\bf x},\tau)\rangle|$ is the modulus of the bulk magnetization for free boundary conditions. It was argued in \cite{vortex} that such a behavior requires $F_{\bf s}({\bf p}_1|{\bf p}_2)$ proportional to 
\EQ
\frac{{\bf p}_1-{\bf p}_2}{|{\bf p}_1-{\bf p}_2|^{3}}\,;
\label{ffs}
\EN
it was shown in the same paper that, when inserted in (\ref{vPhi0}), this expression produces the result 
\bea
\langle{\bf s}({\bf x},0)\rangle_{\cal B} &\sim& v\,\frac{\sqrt{\pi}}{2}\,{}_1F_1\left(\frac{1}{2},2;-\eta^2\right)\eta\,\hat{\bf x}\,,\label{result}\\
&=& v\,\frac{\sqrt{\pi}}{2}\,\eta \left[I_{0}(\eta^{2}/2) + I_{1}(\eta^{2}/2) \right]  e^{-\eta^{2}/2}\,\hat{\bf x}\,,
\nonumber
\eea
where ${}_1F_1(\alpha,\gamma;z)$ is the confluent hypergeometric function, $I_k(z)$ are Bessel functions, and
\EQ
\eta\equiv \sqrt{\frac{2m_V}{R}}\,|{\bf x}|\,.
\label{eta}
\EN
The generalization of (\ref{result}) to $\langle{\bf s}({\bf x},\tau)\rangle_{\cal B}$ is straightforward and results in replacing $\eta$ by $\eta/\sqrt{1-4\tau^2/R^2}$.

We now compare the theoretical prediction (\ref{result}) with Monte Carlo simulations of the $XY$ model on the cubic lattice. Obviously, in the simulations $L$ is finite, but we always take it sufficiently large to exhibit the approach to the theoretical asymptotic values; lattices with $L$ up to 161 and $R$ up to 91 are considered (lengths entering simulations are expressed in units of the lattice spacing). Technical details are quite similar to those of our recent study of two-dimensional Potts models \cite{DSS}. In particular, the standard Metropolis algorithm \cite{LB} is used. Thermal averages are computed by averaging over several (at least six) realizations, with each run being of length $10^6$ Monte Carlo steps per site. The resulting error bars are not depicted in figures 2 and 3 below; usually their size does not exceed that of the symbols in those figures.

\begin{figure}[t]
\begin{center}
\includegraphics[width=8.5cm]{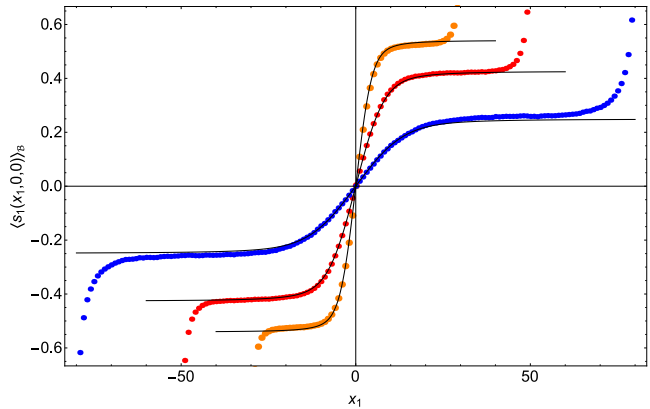}
\caption{Analytic values of the magnetization (\ref{result}) (continuous curves) and corresponding Monte Carlo results (data points). In order of decreasing slope at $x_1=0$, the curves refer to ($T=2.0$, $R=31$), ($T=2.1$, $R=61$), and ($T=2.18$, $R=61$). The Monte Carlo curves were obtained for $L=61$, $101$, $161$, respectively.}
\label{magnetization}
\end{center}
\end{figure}

The relations we write below are intended in the limit in which the temperature approaches the critical value $T_c$, which is known very accurately; we quote here the result $1/T_c=0.4541652(11)$ \cite{CHPV} and take $T_c\simeq 2.2018$. Our simulations are performed in the spontaneously broken phase $T<T_c$, sufficiently close to $T_c$ to make corrections to scaling inessential, at least within the level of accuracy relevant for the purposes of this paper. Since for $L$ large and sufficiently away from the boundaries all radial directions in the plane $\tau=0$ are equivalent, we focus on the cases $x_2=0$ or $x_1=0$. We measured $\langle{\bf s}_i\rangle_{\cal B}$ along these axes and verified that, within error bars, only the radial component is non-zero. This component should then be compared with (\ref{result}), taking into account that near $T_c$
\bea
m_V &\simeq & m_V^0\,(T_c-T)^\nu\,,\\
v &\simeq & v_0\,(T_c-T)^\beta\,,
\eea
with $\nu=0.6717(1)$, $\beta=0.3486(1)$ \cite{CHPV}, and $v_0=0.945(5)$ \cite{EHMS}. It follows that $m_V^0$ is the only unknown quantity in the comparison between theory and data. Our Monte Carlo results for different values of $T$ and $R$ are shown in figure~\ref{magnetization}, and seen to be in remarkable agreement with the theoretical curves corresponding to $m_V^0=2.5$. In particular, the figure confirms that the magnetization profiles depend on the scaling variable (\ref{eta}), which in turn originates from the fact that the vortex is an asymptotic particle of the underlying 	QFT. The data also implicitly confirm the form (\ref{ffs}) of $F_{\bf s}({\bf p}_1|{\bf p}_2)$. A singularity for equal momenta is known to appear also in the form factor of the spin field on the soliton state in two dimensions \cite{BKW,AT}, where it accounts for new results in the theory of phase separation and interfaces \cite{DS}. 

As any critical amplitude, $m_V^0$ is non-universal (i.e. depends on lattice details), and it is relevant to obtain the universal ratio with another mass amplitude. Above $T_c$, the correlation function $\langle{\bf s}(x)\cdot{\bf s}(0)\rangle$ decays exponentially at large distances as $e^{-m_+|x|}$, where $m_+\simeq m_+^0\,(T-T_c)^\nu$ is the mass of the lightest particles and coincides (see e.g. \cite{Kogut}) with the inverse of the correlation length determined numerically in \cite{Hasenbusch}. From the data reported in that paper (see Table~7 and Eq.~(25)) we deduce the value $m_+^0\simeq 1.21$, which then leads to the universal relation (\ref{ratio}).

Below $T_c$, the presence of Goldstone bosons makes more complicated extracting a mass scale from the decay of spin-spin correlations, and it has been common to consider instead the helicity modulus $\Upsilon$, which measures the free energy change under a twist of the spins \cite{FBJ}. On the other hand, we are seeing that a true mass, $m_V$, emerges from the measurement of the magnetization (\ref{result}). Together with (\ref{ratio}), the result $\Upsilon/m_+=0.411(2)$ obtained in \cite{Hasenbusch} then leads to the universal ratio $\Upsilon/m_V\approx 0.2$. 

\begin{figure}[t]
\begin{center}
\includegraphics[width=8.5cm]{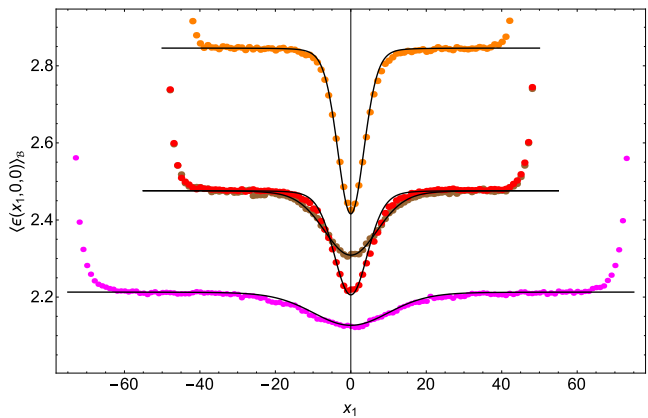}
\caption{Analytic values of the energy density profile (\ref{Geps}) (continuous curves) and corresponding Monte Carlo results (data points). The top and bottom profiles refer to ($T=2.0$, $R=31$) and ($T=2.16$, $R=91$), respectively. The profiles in between are obtained for $T=2.1$ and refer to $R=31$ (deeper minimum) and $R=81$. Simulations were performed with $L=91,101,151$ for $T=2.0,2.1,2.16$, respectively.}
\label{vortex_energy}
\end{center}
\end{figure}

We also measured the local energy density $\varepsilon_i=\sum_{j\sim i}{\bf s}_i\cdot{\bf s}_j$, where the sum runs over the nearest neighbors of site $i$. The Monte Carlo data we obtained along radial directions at $\tau=0$ are shown in figure~\ref{vortex_energy} and clearly exhibit the localization of the vortex energy around the center of the system. They also allow to see that the depth of the minimum of the profiles scales as $R^{-1/2}$. These features are accounted for by the choice $F^c_\varepsilon({\bf p}_1|{\bf p}_2)\propto (|{\bf p}_1||{\bf p}_2|)^{-1/2}$ for the energy density field $\varepsilon(x)\sim{\bf s}^2(x)$; the superscript $c$ denotes the connected part. Upon insertion in (\ref{vPhi0}) this yields
\EQ
\langle\varepsilon({\bf x},0)\rangle_{\cal B}\sim\frac{A}{\sqrt{R}}\left[{}_1F_1\left(\frac{3}{4},1,-\frac{m_V}{R}{\bf x}^2\right)\right]^2+{\cal E}_\textrm{vac}\,,
\label{Geps}
\EN
where the additive constant ${\cal E}_\textrm{vac}=\langle 0|\varepsilon|0\rangle$, corresponding to the vacuum energy density, comes from the disconnected part $(2\pi)^2m_V\delta({\bf p}_1-{\bf p}_2)\langle 0|\varepsilon|0\rangle$ of $F_\varepsilon({\bf p}_1|{\bf p}_2)$. The quantities appearing in (\ref{Geps}) scale as
\bea
{\cal E}_\textrm{vac} &\simeq & {\cal E}_0\,(T_c-T)^{\nu X_\varepsilon}\,,\\
A &\simeq & A_0\,(T_c-T)^{\nu(X_\varepsilon-1/2)}\,,
\label{ampl}
\eea
where $X_\varepsilon$ is the scaling dimension of the energy density field; since $\nu=1/(3-X_\varepsilon)$, the value we already quoted for this exponent yields $X_\varepsilon\simeq 1.51$. For $|{\bf x}|$ large and sufficiently away from the boundary, the Monte Carlo data for the energy density asymptotize to the bulk energy density ${\cal E}$, which differs from ${\cal E}_\textrm{vac}$ by regular terms $c_n(T_c-T)^n$, $n=0,1,\ldots$ (see e.g. \cite{Hasenbusch}); we obtain ${\cal E}$ fitting the data reported in \cite{Hasenbusch} for a list of values of $T$. Having already determined $m_V$, the only unknown parameter left in the comparison between theory and data for the energy density is the amplitude $A_0$ entering (\ref{ampl}). The value $A_0=-7.1$ yields the agreement with the Monte Carlo data exhibited in figure~\ref{vortex_energy}. It is remarkable how the comparison with the simulation data allows to correct the behavior $F_\varepsilon^c({\bf p}_1|{\bf p}_2)=\textrm{constant}$ assumed \footnote{This low-energy behavior for the energy density is known to hold on the soliton state in two dimensions, see  \cite{AT} and F.A. Smirnov, Form Factors in Completely Integrable Models of Quantum Field Theory, World Scientific, 1992.} in \cite{vortex}, and to gain further insight into this previously unexplored sector of QFT. 

It follows from (10) that $|\langle{\bf s}({\bf x},0)\rangle_{\cal B}|/v$ behaves as $A |{\bf x}|$ and $1-(B |{\bf x}|)^{-2}$ for small and large $|{\bf x}|$, respectively, with $A/B=\sqrt{\pi}/4\simeq 0.443$. We notice that these asymptotics are also exhibited by the numerical solution (see \cite{KP}) of the Gross-Pitaevskii (GP) equation for the wave function of a vortex in a Bose gas, with a value $0.412$ for $A/B$. In this form, the comparison overcomes the fact that the characteristic lengths are different in the two cases; in particular, the length $\sqrt{R/2m_V}$ in (\ref{eta}) depends on the distance $R$, which has no counterpart in the GP calculation. In perspective, it will be interesting to see whether our results can be relevant for the controversial problem of defining an inertial mass per unit length of vortex tubes in superfluids (see \cite{inertial}). 

It must be noted that the result (\ref{result}) for $\langle{\bf s}(x)\rangle_{\cal B}$ relies only on the topological constraints, and does not require that the scalar field interacts only with itself. Hence, (\ref{result}) should hold also if the scalar is coupled to the electromagnetic field \footnote{This is the case discussed in textbooks  because it overcomes Derrick's theorem; it is also relevant for superconductivity (see e.g. \cite{Coleman,Weinberg,Ryder}).}. This case, however, does not correspond to the $XY$ universality class and, in particular, the mass ratio (\ref{ratio}) will be different. Similarly, (\ref{result}) should hold in the broken phase of $XY$ models allowing for antiferromagnetic bonds. Vortex lines in a model of this type have been considered \cite{Kusmartsev} in connection with the paramagnetic Meissner effect.

Summarizing, we studied the spontaneously broken phase of the three-dimensional $XY$ model with boundary conditions enforcing the presence of a vortex line. Through comparison with analytic expressions, we showed that the results of Monte Carlo simulations for the order parameter and energy density profiles correspond to a field theory possessing the vortex as a stable quantum particle, and determined in the process the numerical value of its mass. The result also yields the first direct verification that Derrick's theorem, as a statement for classical field configurations, does not provide a fundamental obstruction to the existence of topological particles in purely scalar QFTs in more than two dimensions. The analytic form of the profiles for large end-to-end distance of the vortex line relies on topological properties and should continue to hold when the scalar field is coupled to electromagnetism.


\vspace{.5cm}
 \noindent \textbf{Acknowledgments.} AS thanks the \'Ecole de Physique des Houches
during the session "Integrability in Atomic and Condensed Matter Physics" for hospitality during completion of this work.

\end{document}